\documentclass[twocolumn]{aastex631}

\newcommand{\RomanNumeralCaps}[1]
    {\MakeUppercase{\romannumeral #1}}
\graphicspath{{./}{figures/}}

\begin{document}

\title{Predicting the Yields of  $z>6.5$ Quasar Surveys in the Era of Roman and Rubin}

\author[0000-0003-0747-1780]{Wei Leong Tee}
\affiliation{Steward Observatory, University of Arizona, 
933 N Cherry Ave, Tucson, AZ 85719, USA }

\author[0000-0003-3310-0131]{Xiaohui Fan}
\affiliation{Steward Observatory, University of Arizona, 
933 N Cherry Ave, Tucson, AZ 85719, USA }

\author[0000-0002-7633-431X]{Feige Wang}
\affiliation{Steward Observatory, University of Arizona, 
933 N Cherry Ave, Tucson, AZ 85719, USA }

\author[0000-0001-5287-4242]{Jinyi Yang}
\affiliation{Steward Observatory, University of Arizona, 
933 N Cherry Ave, Tucson, AZ 85719, USA }

\author[0000-0002-9226-5350]{Sangeeta Malhotra}
\affiliation{Astrophysics Science Division, NASA Goddard Space Flight Center, 
8800 Greenbelt Road, Greenbelt, Maryland, 20771, USA}

\author[0000-0002-1501-454X]{James E. Rhoads}
\affiliation{Astrophysics Science Division, NASA Goddard Space Flight Center, 
8800 Greenbelt Road, Greenbelt, Maryland, 20771, USA}



\begin{abstract}

Around 70 $z>6.5$ luminous quasars have been discovered, strongly biased toward the bright end, thus not providing a comprehensive view on quasar abundance beyond cosmic dawn. We present the predicted results of Roman/Rubin high-redshift quasar survey, yielding 3 times more, $2-4$ magnitudes deeper quasar samples, probing high-redshift quasars across broad range of luminosities, especially faint quasars at $L_\mathrm{bol}\sim 10^{10}\;L_{\odot}$ or $M_\mathrm{1450} \sim-22$ that are currently poorly explored. 
We include high-$z$ quasars, galactic dwarfs and low-$z$ compact galaxies with similar colors as quasar candidates. We create mock catalogs based on population models to evaluate selection completeness and efficiency. 
We utilize classical color dropout method in $z$ and $Y$ bands to select primary quasar candidates, followed up with Bayesian selection method to identify quasars. We show that overall selection completeness $> 80\%$ and efficiency $\sim 10\%$ at $6.5<z<9$, with 180 quasars at $z>6.5$, 20 at $z > 7.5$ and 2 at $z > 8.5$. 
The quasar yields depend sensitively on the assumed quasar luminosity shape and redshift evolution. 
Brown dwarf rejection through proper motion up to 50\% can be made for stars brighter than 25 mag, low-$z$ galaxies dominate at fainter magnitude. Our results show that Roman/Rubin are able to discover a statistical sample of the earliest and faintest quasars in the Universe. The new valuable datasets worth follow up studies with James Webb Space Telescope and Extremely Large Telescopes, to determine quasar luminosity function faint end slope and constraint the supermassive black holes growth in the early Universe.
\end{abstract}


\section{Introduction} \label{sec:introduction}
High-redshift quasars are key probes of the early Universe. 
Spectra of $z \gtrsim 6$ quasars provide crucial information regarding the properties of the intergalactic medium (IGM) during the epoch of reionization (EoR). 
Gunn-Peterson (GP) absorption in high resolution quasar spectra can be used to measure the neutral hydrogen content in the IGM and map the history of reionization \citep{Fan2006AJ....132..117F,Becker2015PASA...32...45B}.
Recent quasar studies suggest that the IGM changes rapidly from being highly neutral to ionized in a relatively short period of time at $5.5 < z < 7.5$ \citep[e.g.][]{Banados2018Natur.553..473B,Davies2018ApJ...864..142D,Wang2020ApJ...896...23W,Yang2020ApJ...897L..14Y}. 
However, more sightlines are needed to sample the IGM evolution, especially at $z>7$, to probe whether this transition is homogeneous or highly patchy. 

The highest redshift quasars also directly measure the evolution of early supermassive black holes (SMBHs) and constrain the formation and growth history of the first generation black hole seeds.
Recent observations of quasars at $z > 6$ confirm the existence of massive SMBHs with $\sim 10^{8-10} M_\odot$ 
\citep[e.g.][]{ Mortlock2011Natur.474..616M,Wu2015IAUGA..2251223W,Banados2018Natur.553..473B,Matsuoka2019ApJ...872L...2M,Matsuoka2019ApJ...883..183M,Shen2019ApJ...873...35S,Wang2021ApJ...907L...1W,Yang2020ApJ...897L..14Y,Yang2021ApJ...923..262Y} in the young Universe, with the highest redshift quasars at $z\sim 7.5$ hosting $10^{9} M_\odot$ SMBH. 
Standard SMBH formation models, in which SMBHs grow via Eddington-limited accretion from stellar mass black hole \citep[e.g.][]{Volonteri2010MNRAS.409.1022V,Volonteri2012Sci...337..544V}, fail to explain the existence of these observed SMBHs, given the extreme short growth time. 
Several theoretical approaches have been investigated to explain the formation of observed SMBHs, through the direct collapse of a primordial cloud \citep[e.g.][]{Bromm2003ApJ...596...34B,Begelman2006MNRAS.370..289B,Ferrara2014MNRAS.443.2410F,Habouzit2016MNRAS.463..529H,Schauer2017MNRAS.471.4878S,Dayal2019MNRAS.486.2336D} into massive black hole seeds ($10^{4-6}\; M_\odot$), or by rapid growth of low mass seeds ($10^{2-3}\; M_\odot$) with periods of Eddington and super-Eddington accretion \citep[e.g.][]{Madau2001ApJ...551L..27M,Ohsuga2005ApJ...628..368O,Tanaka2009ApJ...696.1798T,Inayoshi2016MNRAS.459.3738I}, or growth in radiatively inefficient accretion modes \citep[e.g.][]{Trakhtenbrot2017ApJ...836L...1T,Davies2019ApJ...884L..19D}.
Quasars at higher redshift and lower luminosity than those in the current sample will allow stronger tests to these models. 

The discovery of high-$z$ quasars starts at the beginning of this century, mostly using wide-field imaging data from the Sloan Digital Sky Survey 
(SDSS; e.g. \citealt[][]{Fan2001AJ....122.2833F,Fan2006AJ....132..117F}), 
the Canada-France-Hawaii Telescope Legacy Survey (CFHTLS; e.g. \citealt[][]{Willott2009AJ....137.3541W}), 
the Panoramic Survey Telescope and Rapid Response System 1 (Pan-STARRS 1; e.g. \citealt[][]{Banados2016ApJS..227...11B}), 
the United Kingdom Infrared Telescope Infrared Deep Sky Survey (UKIDSS; e.g. \citealt[][]{Banados2018Natur.553..473B}), 
the VISTA Kilo-degree Infrared Galaxy survey (VIKING; e.g. \citealt[][]{Venemans2013ApJ...779...24V}); 
the VLT Survey Telescope ATLAS (VST-ATLAS; e.g. \citealt[][]{Reed2017MNRAS.468.4702R}); 
the Dark Energy Survey (DES; e.g. \citealt[][]{Yang2019AJ....157..236Y,Wang2019ApJ...884...30W}); 
and the Hyper Suprime-Cam (HSC; e.g. \citealt[][]{Matsuoka2018ApJ...869..150M}) on the Subaru telescope.
More than 200 quasars have now been discovered at $z \geq 6$ from different wide-field surveys \citep[see review][]{Fan2022arXiv221206907F}, and the sample size grows rapidly in the last decade due to the combined effort from deep optical and near-infrared (NIR) observations.

The search for quasars at $z \gtrsim 6.5$ requires photometric information from both optical and near-infrared (NIR) bands. 
Quasar flux passing through the high-$z$ IGM leaves the trademark of a distinct Ly$\alpha$ break with flux blueward being completely absorbed, and therefore distinguishable from other types of objects with blue dropout in colors.
Utilizing multi-band all sky survey information over past decade, $\sim$ 70 quasars at $z > 6.5$ have been discovered, including 3 at $z>7.5$ with the highest redshift at 7.6 \citep[e.g.][]{Mortlock2011Natur.474..616M,Banados2018Natur.553..473B,Wang2018ApJ...869L...9W,Wang2021ApJ...907L...1W,Yang2019AJ....157..236Y,Yang2020ApJ...897L..14Y,Yang2020ApJ...904...26Y,Matsuoka2019ApJ...872L...2M,Matsuoka2019ApJ...883..183M}.
However, the current generation of quasar searches is limited to $z<8$, due to a combination of shallow survey depth, insufficient area coverage, and enormous number of contaminants or junk introduced in the flux-limited sample. The contaminants are pickup up with similar NIR colors, majority of them are early type galaxies at $z \sim 1-2$ (ETGs), and Galactic cool stars or brown dwarfs, in particular M stars, L and T dwarfs (MLTs).
The large number of contaminants in quasar candidates has elevated observational challenges in spectroscopic follow up observations. The success rate or efficiency, defined as the ratio of real quasars classified over possible candidates through spectroscopy, are in the order of few percents in present survey depths of $J < 22.5$ for $z\sim 6.5$ quasar searches \citep{Wang2019ApJ...884...30W}. The efficiency at higher redshift and fainter magnitude drops drastically.

The next generation ground based and space based telescopes, such as the Vera C. Rubin Observatory, the Euclid Telescope, the James Webb Space Telescope (JWST), and the Roman Space Telescope, should transform quasar research at the highest redshift. 
Combining deep optical photometry from Rubin Observatory Legacy Survey of Space and Time (LSST) and Euclid Near Infrared Spectrometer and Photometer (NISP) wide-field imaging, \cite{Euclid2019A&A...631A..85E} investigates the quasar detection to Euclid depth. They report prediction of $\sim 250$ quasars at $7 < z <9$, $\sim 25$ quasars at $8 < z < 9$ within Euclid 15000 deg$^{2}$ area, with moderate contamination at $J \sim 23$. Euclid selection however is flux-limited to high redshift  sources at $J<24$.
To maximize the quasar yield and probe full understanding on quasar properties, deeper near-infrared surveys are required to push toward fainter magnitudes at which quasars are more abundant.
Roman and Rubin combination is therefore a natural solution.

We aim to use the multi-band photometry of Roman and Rubin to develop sophisticated search methodology for $z > 6.5$ quasars. 
The current state of the art methods in discovering the Type 1 high-$z$ quasars include color dropout and color-color cuts \citep[e.g.][]{Wang2019ApJ...884...30W}, Bayesian model comparison (BMC; e.g. \citealt[][]{Mortlock2012MNRAS.419..390M}), SED model fitting \citep[e.g.][]{Reed2017MNRAS.468.4702R}, random forest classification \citep[e.g.][]{Schindler2017ApJ...851...13S}, and extreme de-convolution (XD; \citealt{Nanni2022MNRAS.515.3224N}). 
In this paper we study joint selection methods: color dropout and color-color cut, followed with the Bayesian model comparison technique, to optimize the quasar yield and minimize the contamination. The paper is structured as follow: we describe the survey properties, especially the data sets that we are using, and the photometric models in Sec \ref{sec:survey_properties}. We explain the population models and mock catalog generation in Sec \ref{sec:simulation_catalog_construction}. In Sec \ref{sec:high_redshift_quasar_selection} we present the detail of each selection method. In Sec \ref{sec:results} we show the results from the simulation in terms of selection completeness. We apply selections to mock catalogs to justify our selection functions, discuss the limitations and summarize the expected yield. In Sec \ref{sec:discussion} we discuss the effect of proper motion and the benefit of this study. Finally we conclude in Sec \ref{sec:summary}. Through the paper we adopt a flat $\Lambda$CDM cosmology model with $H_\mathrm{0}=70 \; \mathrm{kms}^{-1}\mathrm{Mpc}^{-1}$, $\Omega_\mathrm{m} = 0.272$, and $\Omega_\mathrm{\Lambda} = 0.728$ \citep{Komatsu2009ApJS..180..330K}, following \cite{McGreer2013ApJ...768..105M} and \cite{Yang2016ApJ...829...33Y} where the quasar simulation has been inherited.
All magnitudes and colors used, unless specifically mentioned, are on AB system, while uncertainties quoted are at $1 \, \sigma$ confidence level.

\section{Survey Properties} \label{sec:survey_properties}
\begin{figure}[h]
    \centering
    \includegraphics[width=0.47\textwidth]{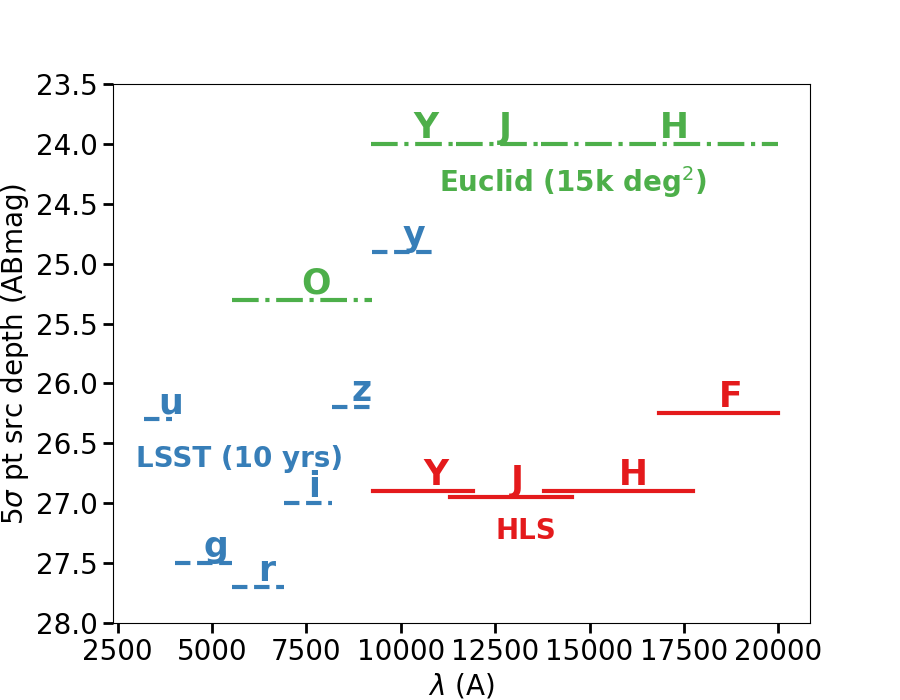}
    \caption{Wavelength coverage and depth of each survey up for the full mission time span. Both Roman HLS and Rubin LSST will reach $5 \,\sigma$ depth of 26 in general. We do not use LSST near-IR $Y$ because it is shallower than Y-band with Roman WFI.}
    \label{fig:survey_sensitivities}
\end{figure}

In this section, we consider photometric properties of Roman and Rubin wide-field surveys.   
The surveys' wavelength coverage and depth are shown in Table \ref{tab:photometry} and Fig \ref{fig:survey_sensitivities}. We introduce additional astrometry information to aid proper motion selection discussed in Sec \ref{sec:proper_motion}. 
The construction of photometric catalog data is presented in Sec \ref{sec:photometric_model}.

\subsection{Roman High Latitude Wide Area Survey (\textit{HLS})}
The Roman HLS consists of both imaging (HLImagingS, HLIS) and low resolution (grism) spectroscopic (HLSpectroscopyS, HLSS) observations over a common survey region using the Wide Field Instrument (WFI) on the Roman Space Telescope,  which has 100 times larger field of view than Wide Field Camera 3 (WFC3) on the Hubble Space Telescope (HST) \citep{WFIRST2012arXiv1208.4012G,WFIRST2015arXiv150303757S}. The survey will map twice covering approximately 2000 deg$^{2}$ in 5 years after the expected launched date of 2026/2027. Survey footprint has been selected to avoid Ecliptic and Galactic plane, and overlapped with optical ground telescopes. Current survey region has been placed near to ra = [0,47.5] deg, dec = [-47.5,0] deg, or $\sin b\sim 1$.
HLIS will reach $\sim 26-27$ ($5\,\sigma$ point source) in four NIR bands: $Y,J,H,F$. 

At the time of this writing, the survey design is not yet finalized and the observing strategy may continue to evolve. For example, the possibility of adding a redder $K_\mathrm{s}$ filter ($2.0-2.4\;\mu$m) has been discussed in \cite{Stauffer2018arXiv180600554S}, with which quasars and galaxies can be better separated from stars with using $Y-K_\mathrm{s}$ color than using $Y-H$ alone. We do not include the additional $K_\mathrm{s}$ filter in this work.
The final observing program details and cadence will be defined by future community process. 
In this work we only use information based on HLIS imaging for quasar search, adopting the $YJHF$ filters for our photometric measurements, and we assign random sky coordinates for sources drawn from HLIS survey region.

\subsection{Rubin Legacy Survey of Space and Time (\textit{LSST})}
Quasars at high-$z$ are expected to have negligible flux blueward of redshifted Ly$\alpha$. 
Deep optical data are needed to characterize the strong Lyman break.  
Possible crossover surveys that meet the criteria are the 6 years space-based Euclid Wide Survey and 10 years ground based LSST. 
Euclid Wide Survey \citep{Euclid2022A&A...662A.112E} has $10 \,\sigma$ depth at 24.5 in $O$ band, which is an optical wide filter spanning 4900-9300 \AA. 
LSST has $i,z,y$-bands reaching $5\,\sigma$ depth at 26.9, 26.1 and 24.9 at the end of 10 years, assuming uniform depth across the survey region. 
LSST broad bands provide sufficient sensitivity and contrast for color dropout selection. We consider LSST as the main source for optical data.

\subsection{Astrometry}\label{sec:astrometry}
Cool dwarfs with high proper motions can be distinguished from stationary quasars using imaging surveys with multiple epochs data.
Various studies have been carried out to characterize the kinematics of cool brown dwarfs in Galactic disk and halo \citep{Faherty2009AJ....137....1F,Best2018ApJS..234....1B}. 
We can cross-validate target positional information using forced photometry to identify moving cool dwarfs. 
Roman WFI has pixel size and full width half maximum (FWHM) for point-spread function (PSF) of $0\arcsec.1$ and $0\arcsec.27$ in $H$ filter \citep{Troxel2023MNRAS.tmp..717T}. The expected $5 \,\sigma$ detection in proper motion is $4-16 \; \mathrm{mas}\;\mathrm{year}^{-1}$, depending on the timing of Roman second pass imaging \citep{WFIRST2019JATIS...5d4005W}.
Rubin LSST maps its survey region with 20 observations each year. Estimated astrometry accuracy is $\sim 50$ mas for sufficiently bright objects at the end of the ten-year survey \citep{Ivezic2019ApJ...873..111I}. To estimate the performance of Rubin in measuring the cool dwarf proper motion, 
we query rubin\_sim\footnote{\href{https://github.com/lsst/rubin_sim}{https://github.com/lsst/rubin\_sim}}, a simulation tool of Rubin Observatory, 
look for all detected LT cool dwarfs (primary spectral type of dwarf contaminants for high-redshift quasar search) over a 10 years baseline.
We obtain the distances of the population, and with an assumed tangential velocity $30 \, \mathrm{km}\,\mathrm{s}^{-1}$ (see Sec \ref{sec:mlt_modelling}), we derive their respective proper motions. 
The furthest cool dwarf marks the lower bound of proper motion measured by Rubin, and it is a function of spectral type.
The median of minimum proper motion detectable by Rubin across spectral types is found to be 30 $\mathrm{mas}\;\mathrm{year}^{-1}$.

\subsection{Advantages of Roman and Rubin}
LSST is scheduled to start science operation no latter than 2024. By 2028 Roman will have collected 1 year imaging data while LSST will have at least 4 years of imaging data within HLIS survey area.
Euclid and Roman are best-suited for different types of high-$z$ quasar science: Euclid eventually will map much larger sky area for the discoveries of luminous quasars, which is crucial for studying massive SMBHs growth in the early Universe and the history of reionization;  
Roman will discover more low luminosity quasars, with bolometric luminosity close to that of typical AGN ($L_\mathrm{bol}\sim 10^{10}\, L_{\odot}$), which enable the studies of evolution and abundance of SMBHs in the early Universe.

\subsection{Photometric Model}\label{sec:photometric_model}
Simulated photometric data with realistic noise that matches the expected observed datasets is necessary for investigating selection efficiency and completeness. 
\cite{Troxel2023MNRAS.tmp..717T} reports that the dominant noises are thermal background, Poisson noise and readout noise in Roman detectors performance with Rubin/Roman imaging simulation. Our primary interests are quasars near the IR detecting limit, which are usually background-limited sources. 
After the noiseless photometry has been generated by integrating the simulated spectrum in Sec \ref{sec:simulation_catalog_construction} over the respective filter throughput, the observed photometric errors are added as Gaussian noise with a lower bound of 0.01 for systematic error. We also include an additional $10 \%$ scatter for the construction of observed photometry to match with observational magnitude error trend. We list the photometry information used in this work in Table \ref{tab:photometry}.

\begin{deluxetable}{cccc}
\tablenum{1}
\tablecaption{Photometric information used in this work.}
\label{tab:photometry}
\tablewidth{0pt}
\tabletypesize{\scriptsize}
\tablehead{
\colhead{Surveys} & \colhead{Band} &\colhead{$\lambda$ (\AA)} &\colhead{Depth ($5\,\sigma$)} 
}
\decimalcolnumbers
\startdata
\begin{tabular}{c}
     Roman  \\
     HLS (5 yrs) 
\end{tabular} &
\begin{tabular}{c}
     Y  \\ J \\ H \\ F
\end{tabular} & 
\begin{tabular}{c}
     9270 - 11920  \\ 11310 - 14540 \\ 13800 - 17740 \\ 16830 - 20000
\end{tabular} &  
\begin{tabular}{c}
     26.9  \\ 26.95 \\ 26.9 \\ 26.25
\end{tabular} 
\\
\begin{tabular}{c}
     Rubin  \\
     LSST (10 yrs) 
\end{tabular} &
$z$ & 8030 - 9380 &26.2 
\enddata
\end{deluxetable}

\section{Simulation \& Catalog Construction} \label{sec:simulation_catalog_construction}
There are two different kind of simulated catalogs used in this study. We first generate a complete set of quasars distributed uniformly in a $(J,z)$ grid. Simulation details are presented in Sec \ref{sec:simulation_of_quasar_selection_function}. This quasar simulation is used to determine the optimum color dropout and color-color cut selection. We then describe the population modelling, mainly on the number densities and color templates of population as a function of magnitude or redshift in Sec \ref{sec:population_models}. Sec \ref{sec:mock_catalog_construction} describes the process of mock catalogs construction to enable realistic estimation of the completeness and efficiency of quasar selection.

\subsection{Simulation of Quasars for Selection Function}\label{sec:simulation_of_quasar_selection_function}
Construction of color selection function requires a complete set of simulated quasars matched with observation. \cite{Fan1999AJ....117.2528F} proposes an empirical method to construct quasar spectrum using observed continuum slope and emission line strengths, assuming shape of quasar spectral energy distributions (SEDs) does not evolve with redshift, as well as models of intergalactic absorption blueward of Ly$\alpha$ emission. This approach has been extended to include higher redshift quasars in a number of recent works \citep[e.g.][]{McGreer2013ApJ...768..105M,McGreer2018AJ....155..131M,Yang2016ApJ...829...33Y,Jiang2016ApJ...833..222J,Wang2019ApJ...884...30W}. 
In this section we briefly describe the spectral model and extend the simulation to include $z=6-10$ quasars. 
The simulated quasar continuum spectrum is assumed to be a broken power law continuum $f_\nu \propto \nu^{-\mu}$ with breaks at 1100, 5700, 9730, 23820, and 30000 \AA. 
The slopes in between these break points are chosen from Gaussian distributions with $\mu = -0.44, -0.48, -1.74, -1.17$ and $\sigma = 0.3$. Various emission lines, including Ly$\alpha$ and C$\mathrm{\RomanNumeralCaps{4}}$, and FeII lines \citep{Boroson1992ApJS...80..109B,Vestergaard2001ApJS..134....1V,Gilkman2006ApJ...640..579G,Tsuzuki2006ApJ...650...57T} are added to the continuum spectrum as Gaussian profiles.
The distribution of emission line features, such as the continuum spectral slope, line equivalent width (EW), and line FWHM are designed to match the colors of SDSS BOSS quasars in the redshift range of $2.2<z<3.5$ \citep{Ross2012ApJS..199....3R}. Quasars are generated via uniform grid of ($J,z$), where $18<J<28$ with $\Delta J = 0.1$ and $6<z<10$ with $\Delta z=0.1$; there are $100$ quasars at each ($J,z$) grid position, with a total of 404,000 quasars generated with absolute magnitude at 1450 \AA, $M_\mathrm{1450}$, ranges from -30 to -18. The simulated spectra correctly reproduce quasar spectral characteristics seen in observation, e.g. Baldwin effect \citep{Baldwin1977ApJ...214..679B} and blueshifted lines \citep{Gaskell1982ApJ...263...79G,Richards2011AJ....141..167R}.

\subsection{Population Models}\label{sec:population_models}

\begin{figure}
    \centering
    \includegraphics[width = \linewidth]{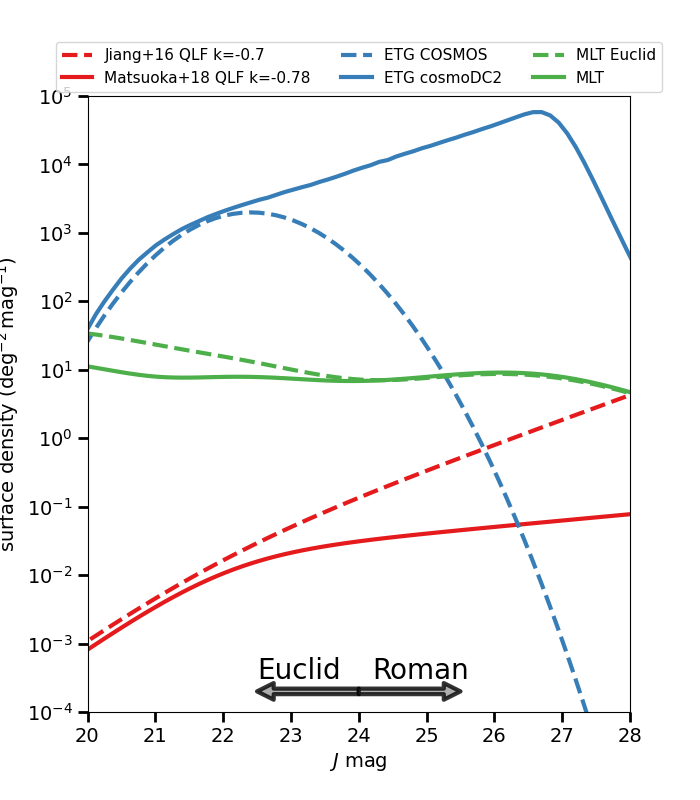}
    \caption{{\bf Candidate population surface densities as a function of J band magnitude. Detailed models can be found in Sec \ref{sec:population_models}. The vast difference in quasar and galaxy densities in faint magnitudes points out the distinct target strategy between \cite{Barnett2021MNRAS.501.1663B} (see Fig 3 in their work) and this work.}\\
    \textit{Red dashed:} Quasars, by extrapolating the luminosity function and spatial density decline rate in \cite{Euclid2019A&A...631A..85E}, and integrating over redshift z = 6.5 - 9.0.
    \textit{Red solid}: Quasars using parameters from \citetalias{Wang2019ApJ...884...30W} and \citetalias{Matsuoka2018ApJ...869..150M}, integrate over redshift z = 6.5 - 9.0.
    \textit{Blue dashed}: COSMOS quiescent-flagged galaxies at redshift z = 1-2.
    \textit{Blue solid}: cosmoDC2 red sequence galaxies at redshift z = 1-2.
    \textit{Green dashed}: MLTs summed over all spectral types (M0-T8) modeled in \cite{Euclid2019A&A...631A..85E}.
    \textit{Green solid}: MLTs summed over all spectral types in this work (M0-T9) with slight difference in early M population densities.
    }
    \label{fig:surface_densities}
\end{figure}

\subsubsection{Quasars}\label{sec:quasar_modelling}
The quasar luminosity function can be generally parametrized as a double power-law \citep[e.g.][]{Boyle1988MNRAS.235..935B}:
\begin{equation}
\resizebox{0.9\hsize}{!}{$\Phi(M_\mathrm{1450},z) = \frac{\Phi^{\ast}(z)}{10^{0.4(\alpha+1)(M_\mathrm{1450}-M^{\ast}_\mathrm{1450})}+10^{0.4(\beta+1)(M_\mathrm{1450}-M^{\ast}_\mathrm{1450})}}$}
\label{eq:DPLQLF}
\end{equation}
and the spatial density decline rate $k$ is embedded within $\Phi^{\ast}(z)$,
\begin{equation}
\Phi^{\ast}(z) = \Phi^{\ast}(z=6) \times 10^{k(z-6)}
\label{eq:Phi_star_z}
\end{equation}
where $M_\mathrm{1450}$ is the absolute magnitude at 1450 \AA, $\alpha$ and $\beta$ are the faint end and bright end slopes, respectively, $M^{\ast}_\mathrm{1450}$ is the characteristic absolute magnitude or break magnitude measured at 1450 \AA 
 (assume no redshift evolution), and $\Phi^{\ast}$ is the normalization of the LF.
With high completeness quasar samples, QLF has been  measured at $z \sim 5$ and extended to $z=6$ across wide luminosity ranges ($-30 \lesssim M_\mathrm{1450} \lesssim -23$), with $(\alpha,\beta,M^{\ast}_\mathrm{1450}) \sim (-2.0,-3.0,-27)$ \citep[e.g.][]{Willott2010AJ....139..906W,McGreer2013ApJ...768..105M,McGreer2018AJ....155..131M,Jiang2016ApJ...833..222J}.
Previous studies of SDSS quasars at $z>3.5$ show that $k=-0.47$ and this evolution has been extrapolated to predict higher redshift quasars yield \citep[e.g.][]{Fan2001AJ....121...54F,Fan2001AJ....122.2833F,Jiang2008AJ....135.1057J,Willott2010AJ....139..906W}. 
Using deeper data from SDSS Stripe 82 region quasars, \cite{McGreer2013ApJ...768..105M} work shows $k$ redshift evolution between $4 < z < 6$, and proposes $k=-0.7$ at $z=5-6$.

However, the measurement of QLF at $z \geq 5$ remains highly uncertain because of the small sample sizes, especially at  fainter magnitudes \citep[e.g.][]{Willott2010AJ....139..906W,McGreer2013ApJ...768..105M,McGreer2018AJ....155..131M,Venemans2013ApJ...779...24V,Venemans2015ApJ...801L..11V,Kashikawa2015ApJ...798...28K,Yang2016ApJ...829...33Y,Jiang2016ApJ...833..222J,Wang2019ApJ...884...30W}.
\citet[hereafter Jiang+16]{Jiang2016ApJ...833..222J} measures the QLF parameters to be ($\alpha,\beta,M^{\ast}_\mathrm{1450},k$) = (-1.90,-2.8,-25.2,-0.72) using a complete sample of 54 
quasars at $5.7<z<6.4$. Their result of $k=-0.72 \pm 0.11$ suggests that the quasar spatial density drops rapidly after $z=5$. 
Recently \citet[hereafter Matsuoka+18]{Matsuoka2018ApJ...869..150M} measures the QLF at $5.7 < z < 6.5$ with 110 quasars, including the largest low luminosity quasars in the Subaru High-z Exploration of Low-Luminosity Quasars project (SHELLQs; \citealt{Matsuoka2016ApJ...828...26M}). They find a clear break at $M^{\ast}_\mathrm{1450}=-24.90$ and a much flatter faint end slope $\alpha = -1.23^{+0.44}_{-0.34}$, compared with \citetalias{Jiang2016ApJ...833..222J} $\alpha = -1.90^{+0.58}_{-0.44}$, although still within $2\,\sigma$ confidence interval.
\citet[hereafter Wang+19]{Wang2019ApJ...884...30W} use a complete bright quasar samples at $6.4 < z< 6.9$ and find $k=-0.78 \pm 0.18$ at $z=6.0-6.7$, further strengthen the claim of a rapid declination in quasar spatial density from $z=3.0-5.0$ to $z>6.0$.

We adopt the latest \citetalias{Matsuoka2018ApJ...869..150M} QLF parameters and \citetalias{Wang2019ApJ...884...30W} decline rate in our QLF-sampled mock catalog construction.
We fix the parameters as $\alpha=-1.23$, $\beta=-2.73$, $M^{\ast}_\mathrm{1450} = -24.90$, $ \Phi^{\ast} = 10.9 \times 10^{-9} \; \mathrm{Gpc}^{-3}\,\mathrm{mag}^{-1}, k=-0.78$. 

\subsubsection{Early Type Galaxies}\label{sec:galaxy_modelling}
cosmoDC2 is a synthetic galaxy catalog \citep{Korytov2019ApJS..245...26K} produced for the LSST second data challenge (DC2) by \cite{LSST2021ApJS..253...31L}. The catalog is based on cosmological N-body simulation matching up with empirical and semi-analytic galaxy models.
It covers $440 \;\mathrm{deg}^{2}$ of sky area within redshift $0 < z < 3$.  
The catalog contains detailed galaxy properties including redshift, luminosity, stellar mass, size and shape, as well as host halo information and lensing-related parameters.

Quasar selection based on color-color cuts depends primarily on color dropout in the bluer bands.
Red, compact, low-$z$ galaxies (early type or quiescent galaxies; ETGs) detected in optical-infrared bands could be misclassified as quasar candidates. 
Galaxies show tendency in increasing compactness at higher redshift (e.g. \cite{VanDerWel2014ApJ...788...28V}, and will appear unresolved especially at low signal-to-noise (S/N) level. 
\cite{Euclid2019A&A...631A..85E} investigates the degree of contamination by unresolved $z=1.0-2.0$ ETGs for Euclid quasar search. 
They use the quiescent galaxies from COSMOS2015 catalog in \cite{Laigle2016ApJS..224...24L}, which are selected by the rest frame NUV/optical-optical/NIR colour cut, or similarly by the rest-frame $UVJ$ color cut.
They simulate the Euclid Near Infrared Spectrometer and Photometer (NISP) instrument detection and observed photometry within $1\arcsec$ aperture on COSMOS ETGs, and conclude that galaxies with $J>22.0$ may be mistaken as unresolved point sources (see Fig 6 in \citealt{Euclid2019A&A...631A..85E}) due to their small sizes. 
Both Euclid and WFI have similar pixel resolution of $0\arcsec.1$ in NIR bands, we expect the galaxy size information would not contribute significantly in discriminating low S/N quasars and galaxies in our work. We inspect the cosmoDC2 galaxies half-light radii and find that faint galaxies being misidentified as quasar with our selection criteria ($J\sim 26$ with S/N $\sim 15$) indeed have radii $< 0\arcsec.1$. 

cosmoDC2 catalog simulates both star-forming and red sequence galaxies populations, with rest frame colors matching observational datasets. We extract the $z=1-2$ red sequence galaxies and treat them as primary galaxy contaminators ETG in this work.
The cosmoDC2 ETG population density is determined by counting the ETG galaxy number in $(J,z)$ bins within redshift space $z=1-2$ and is shown in Fig \ref{fig:surface_densities}. Densities of cosmoDC2 ETG and COSMOS ETG match well in the brighter magnitudes, while the lack of COSMOS ETG at $J > 23$ leads to underestimation of the density of ETG (note that $J=24$ is the Euclid $5 \,\sigma$ point source magnitude) at fainter magnitude.

\begin{figure*}[ht]
    \centering
    \includegraphics[width=0.9\textwidth]{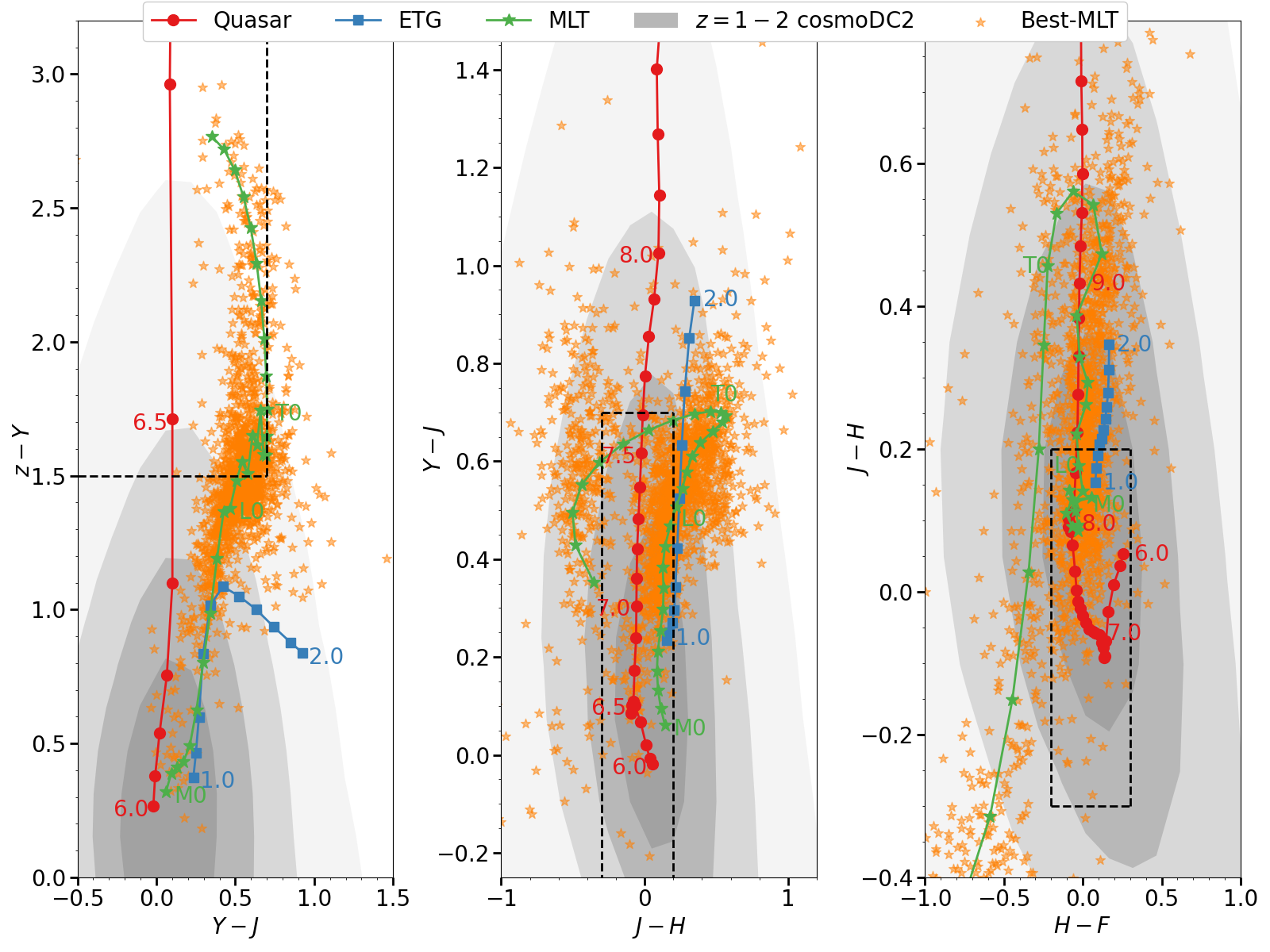}
    \caption{Color-color diagram for the  $z=6.5-7.5$ quasar color selection. Detail modelling and full color selections can be found in Sec \ref{sec:population_models} and Sec \ref{sec:color_dropout_selection_method}. 
    The red tracks are quasar model colors based on the simulation described in Sec \ref{sec:color_dropout_selection_method}, red circles for a redshift grid with $\Delta z=0.1$ spacing.
    The blue squares are median cosmoDC2 ETG colors with $z=1.0,2.0$ labelled, grey contours are the entire galaxy population with 4 iso-proportions of densities from 0.2-0.8 (limited to $J=25.5$ since fainter galaxies are excluded as discussed in Sec \ref{sec:res_color_selection}). 
    The green stars are the MLT model colors with M0, L0 and T0 labelled. We also show orange points of spectroscopic confirmed MLTs in \cite{best_william_m_j_2020_4570814}.\\
    \textit{Left}: $zYJ$ colors for LSST and WFI. Quasars at $z>6.5$ have their Ly$\alpha$ line gradually shifted into longer wavelength infrared bands, leaving them almost undetectable flux in the $z$ band, increasing their $z-Y$ colors with redshifts. The majority of non-quasar population have bluer colors on vertical axis. 
    \textit{Middle}: $YJH$ colors for LSST and WFI. At $z>8.5$ Ly$\alpha$ line shifts from $Y$ to $J$, resulting in redder $Y-J$ colors. From the color tracks it is obvious that $z>8.0$ quasars suffer less contamination from other populations.
    \textit{Right}: $JHF$ colors for LSST and WFI. Note that the late T population (T6 and later) generally have bluer $JHF$ colors and are easily excluded using this color plot.
    }
    \label{fig:population_color_color_track}
\end{figure*}

\subsubsection{Dwarfs (MLTs)}\label{sec:mlt_modelling}
Low mass stars and brown dwarfs with low surface temperature, including late types M,  L and T spectral types,  have red infrared colors, and are known as one of the major contaminant populations for high-$z$ quasar selection in infrared bands from past wide surveys \citep[e.g.][]{Mortlock2012MNRAS.419..390M,Wang2016ApJ...819...24W}.
The majority of MLTs lie in the Galactic thin disk with lower scale height, which are considered to be a young ($\sim 3$ Gyrs) and slow moving ($v_\mathrm{tan} \sim 30 \,\mathrm{kms}^{-1}$) population \citep{Faherty2009AJ....137....1F,Ferguson2017ApJ...843..141F}. 

The procedure to obtain number density is similar to that described in \cite{Euclid2019A&A...631A..85E}; we briefly address the main process and refer readers to their paper for more detail. The number density of the thin disk MLT population can be modeled as $\rho = \rho_\mathrm{0} e^{-Z/Z_\mathrm{s}}$, where $\rho_\mathrm{0}$ is the number density of MLT spectral type at the Galactic central plane, $Z$ is the vertical distance from the plane, and $Z_\mathrm{s}$ is the scale height, assumed to be 300 pc \citep{Gilmore1983MNRAS.202.1025G,Kilic2017ApJ...837..162K} 
Given the Roman HLS location in the high galactic latitude region ($\mathrm{\sin}\,b \sim 1$) with a comparatively small survey area, the dependence of number density on Galactic latitude should be negligible for the purpose of quasar selection. 
We derive the density value from luminosity function and photometry information in \cite{Bochanski2010AJ....139.2679B}, \cite{Dupuy2012ApJS..201...19D} and \cite{Skrzypek2016A&A...589A..49S}. We construct the color template based on a dwarfs catalog \citep{best_william_m_j_2020_4570814} compiled from several whole sky surveys observations, the population model is consistent with \cite{Euclid2019A&A...631A..85E}.

Astrometric information of MLTs are obtained from the UltracoolSheet catalog \citep{best_william_m_j_2020_4570814}. They provide ra, dec, parallax and distance measurements for spectroscopic confirmed MLTs. \cite{Best2021AJ....161...42B} additionally identify a complete MLT sample within 20 pc. 
We derive the medians of the tangential velocity $v_\mathrm{tan}$, total velocity $v_\mathrm{tot}$, and velocity dispersion $\sigma_v$ for each spectral type [Spt-0.5, Spt+0.5] in $\mathrm{kms}^{-1}$, for the complete sample in 20 pc and the entire dataset, respectively. 
We find that the result are not significantly different. 
$(v_\mathrm{tan},\sigma_v)=(30,23)$ and $(v_\mathrm{tot},\sigma_v)=(49,35)$ are broadly in good agreement with \cite{Faherty2009AJ....137....1F} and \cite{Best2018ApJS..234....1B}. 
By assuming no evolution of velocity distribution between bright and faint dwarfs, we sample the MLTs total velocity as a Gaussian distribution on top of the Milky Way rotation curve generated by python package galpy\footnote{\href{https://github.com/jobovy/galpy}{https://github.com/jobovy/galpy}}\citep{Bovy2015ApJS..216...29B}. 
We compute distance modulus for each target and measure their distances and proper motions.

\subsection{Mock Catalog Construction}\label{sec:mock_catalog_construction}
To simulate the realistic survey data and the impact of contaminants on quasar selection, we produce photometric mock catalogs for each population, both luminosity- and spatial-sampled, reaching 1 mag fainter than the survey limit, to allow possible scatter above the S/N limit.

We generate a QLF-sampled quasar catalog (detail in Sec \ref{sec:quasar_modelling}) with the parameters in Eq (\ref{eq:DPLQLF})-(\ref{eq:Phi_star_z}). The number of quasars generated are 10 times larger than the number in 2000 $\mathrm{deg}^{2}$ to minimize the impact of small number statistics. 
The cosmoDC2 catalog contains rest-frame SED (100 nm to 2000 nm) for individual galaxy. We measure the corresponding synthetic WFI magnitudes by integrating against the filter throughput. Note that in constructing the galaxy contaminants we do not explicitly requires the object to be an early type galaxy. The entire galaxy catalog is used to include all possible sources. 
The MLT catalog is generated with their number densities matching the distributions in Fig \ref{fig:surface_densities}. SpeX Prism Library (SPL) and SpeX Prism Library Analysis Toolkit (SPLAT) \citep{Burgasser2017ASInC..14....7B} provide over 3000 low-resolution, near-infrared MLT spectra with spectral class classification.
We construct the composite MLT spectrum by inspecting the suitable MLT spectra, and derive the synthetic magnitudes based on the median colors in each band with respect to $J$.
Observed photometric catalog with multiple realizations are created to minimize possible scatter across experiments.
In all, our mock catalog contains $4 \times 10^3$ WFI-$J$ $5 \,\sigma$ detected quasars, $9 \times 10^5$ MLTs and $4.2 \times 10^9$ galaxies ($20 \%$ ETG).

\section{High-Redshift Quasar Selection} \label{sec:high_redshift_quasar_selection}
\subsection{Color Dropout Selection Method}\label{sec:color_dropout_selection_method}
{\footnotesize
\begin{align}\label{eq:color-color dropout}
    &\mathrm{S/N} (u,g,r)<5.0  \nonumber \\
    &\mathrm{S/N}(J,H,F)>5.0 \nonumber \\
    \text{For $6.5<z<7.5$:}&
    \begin{cases}
    \mathrm{S/N}(Y)>5.0 \\
     \mathrm{S/N}(i)<3.0 \text{ or } i-Y > 3.5 \\
     \mathrm{S/N}(z)<3.0 \text{ or } z-Y > 1.5 \\
     Y-J<0.7  \\
     -0.3 <J-H < 0.2 \\
     -0.2 <H-F < 0.3 \\
    \end{cases} \nonumber \\
    \text{For $7.5<z<8.5$:}&
    \begin{cases}
    \mathrm{S/N}(i)<3.0 \\
     \mathrm{S/N}(z)<3.0 \text{ or } z-Y > 3.5 \\
     \mathrm{S/N}(Y)<3.0 \text{ or } 0.6<Y-J<2.0  \\
     -0.2 <J-H < 0.3 \\
     -0.3 <H-F < 0.1 
    \end{cases} \nonumber \\
     \text{For $8.5<z<9.0$:}&
     \begin{cases}
     \mathrm{S/N}(i,z)<3.0 \\
     \mathrm{S/N}(Y)<3.0 \text{ or } Y-J>1.5  \\
     0.1 <J-H < 0.6 \\
     -0.2 <H-F < 0.2 
    \end{cases}
\end{align}}%

To select quasar candidates, we first require  $5 \,\sigma$ S/N non-detection in optical $u,g,r$ and $5 \,\sigma$ S/N detection in $J,H,F$. 
In literature works \citep[e.g.][]{McGreer2018AJ....155..131M} S/N of blue dropouts are generally set to $<3 \,\sigma$, we initially use $5 \,\sigma$ to avoid missing quasar candidates, follow up investigation on real quasar candidates reveals that they also satisfy S/N$(u,g,r)<3\,\sigma$ criteria. We do not modify the dropout S/N to conserve the high completeness in selecting quasar candidates. 
Also note that $F$ band $5 \,\sigma$ limit is shallower than those in other bands. As a result, we are actually requiring S/N $\sim 10$ in J band for any source to be selected.
We split the color-color cut selection into three redshift ranges: $6.5<z<7.5$, $7.5<z<8.5$, $8.5<z<9.0$. The regions are characterized by optical-NIR drop out in $i-Y,\;z-Y$ and infrared drop out $Y-J$.
The regions of $J-H$ and $H-F$ are chosen to optimize the quasar selection and minimize the contaminants cross-over. We summarize the color dropout and color-color cut criteria in Eq \ref{eq:color-color dropout} and Fig \ref{fig:population_color_color_track}. 

\subsection{Bayesian Model Comparison}
\cite{Mortlock2012MNRAS.419..390M} propose a Bayesian-based method to discover high-redshift quasars. The main idea is to calculate a posterior quasar probability, $P_\mathrm{q}$, for a target to be quasar rather than others. In this work we consider three closely related population in color space, namely quasars ($q$); early type galaxies ($g$); and MLTs ($s$). $P_\mathrm{q}$ can be expressed as,
\begin{equation}
    P_\mathrm{q} \equiv P(q \vert \bm{d}) = \frac{W_\mathrm{q}(\bm{d})}{W_\mathrm{q}(\bm{d}) + W_\mathrm{g}(\bm{d}) +W_\mathrm{s}(\bm{d})}
\end{equation}
where $\bm{d}$ is the photometric data.

To calculate the weights of each population for a given source, we make use of the available photometric fluxes and uncertainties, combined with the surface density model of the population as a prior. For every single source, assume it belongs to a particular population, we compute a Gaussian likelihood function based on model fluxes, 
\begin{equation}
    p(\bm{d} \vert \theta_i,i) = \prod_{b=1}^{N_\mathrm{b}} \frac{1}{\sqrt{2 \pi} \hat{\sigma}_b} \mathrm{exp} \Bigg\{ -\frac{1}{2}\left[\frac{\hat{f_b}-f_b(\bm{\theta_i})}{\hat{\sigma}_b}\right]^2 \Bigg\}
\end{equation}
where $b$ is the photometric band, $\hat{f_b}$ and $\hat{\sigma}_b$ are the observed flux and error, $f_b(\bm{\theta_i})$ is the model flux, $\theta_i$ is the model parameters govern the population color, $i=q, g, s$ respectively. 
Note that current formalism only support scatter in brown dwarf colors, as seen in some studies \citep[e.g.][]{Skrzypek2016A&A...589A..49S}. For simplicity we assume no scatter in population model colors, the generalization of including scattering of population model color will be considered in future works.
We combine the Gaussian likelihood function with the surface density function $\Sigma_t$, integrate over parameter spaces to obtain corresponding weights, 
\begin{equation}
W_i(\bm{d}) = \displaystyle \int \Sigma_t (\theta_i) \; p(\bm{d} \vert \theta_i,i) \;\;d\theta_i
\end{equation}
in order to calculate the final posterior quasar probability. 
In \cite{Mortlock2012MNRAS.419..390M} $P_\mathrm{q}=0.1$ was used as the threshold to select quasar candidates. The value is chosen to balance the quasar selection completeness and contamination rate, and proved to be an effective cut in UKIDSS LAS and other high-redshift quasar surveys \citep{Mortlock2012MNRAS.419..390M,Barnett2021MNRAS.501.1663B}, where changing $P_\mathrm{q} = 0.07-0.2$ does not greatly impact the overall selection \citep[e.g.][]{Barnett2021MNRAS.501.1663B,Nanni2022MNRAS.515.3224N}.
In this work we adopt similar value of $P_\mathrm{q}=0.1$ for quasar candidate selection.

\section{Results} \label{sec:results}
Our goal is to investigate the potential of joint color selection and BMC. We first apply color selection, then follow up with Bayesian probability selection to the mock catalogs generated to identify high-$z$ quasars. We discuss the expected yield, completeness and efficiency of color selection method in Sec \ref{sec:res_color_selection}. Improvement over Bayesian method is addressed in Sec \ref{sec:res_bmc_selection}.

\begin{figure*}[!b]
    \centering
    \includegraphics[width=\textwidth]{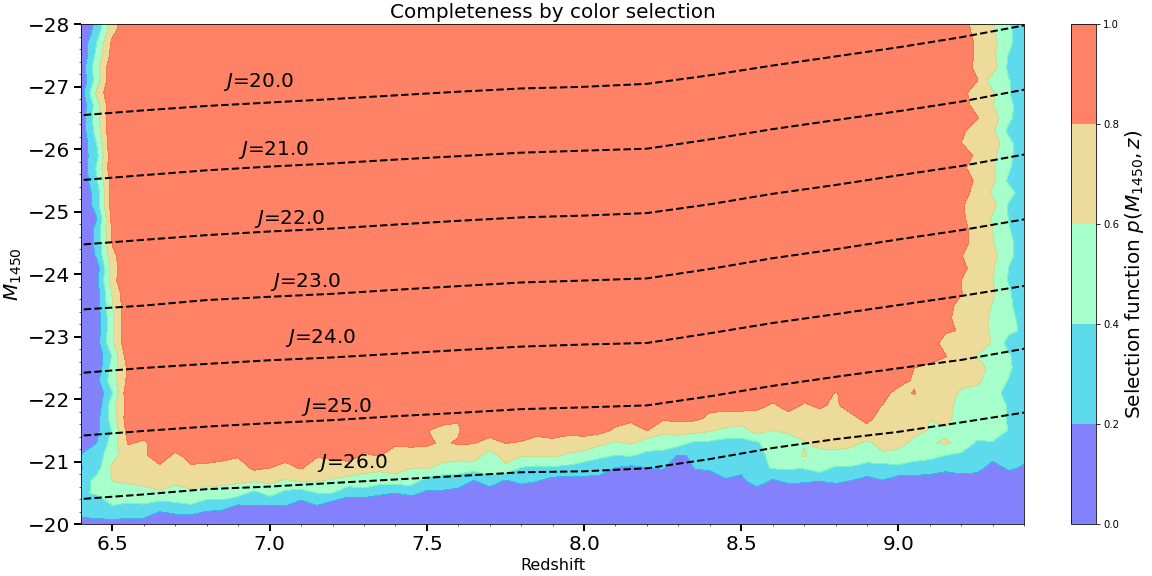}
    \caption{Completeness of color selection function of $6.5<z<9.0$ quasars. The probability is the fraction of simulated quasars selected by color selection method among all simulated quasars in the ($M_\mathrm{1450},z$) bin. Black dashed line is the $J$ band magnitude across the redshift range. Color selection is capable to select quasar down to $J \sim 25.5$ at all redshifts. 
    The selection function is mostly complete at $J< 25.5$ (S/N$>20$).}
    \label{fig:completeness_M1450z_colorcut_z65_90}
\end{figure*}

\begin{figure*}[ht]
    \centering
    \begin{minipage}[t]{.49\textwidth}
    \centering
    \includegraphics[width =  \linewidth]{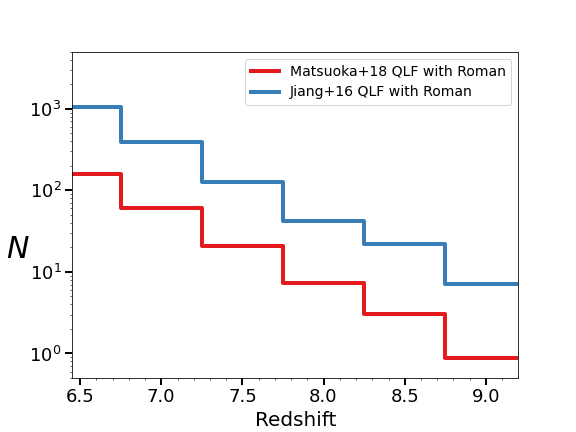}
    \end{minipage}
    \unskip\hfill
    \begin{minipage}[t]{.49\textwidth}
    \centering
    \includegraphics[width = \linewidth]{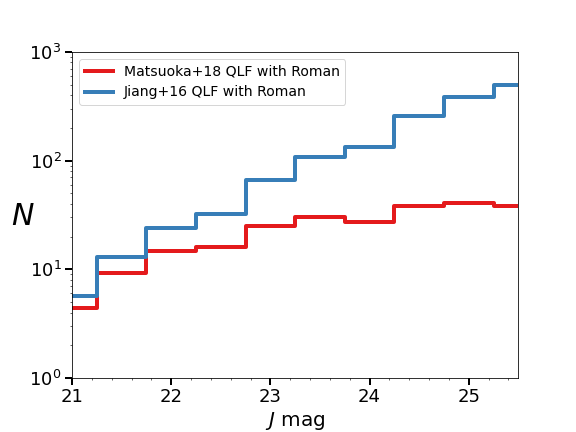}
    \end{minipage}
    \caption{\textit{Left}: Predicted quasar yield in 0.5 redshift bin, \textit{Right}: Predicted quasar yield in 0.5 magnitude bin.\\
    Results are derived from color cut selection in a 2000 deg$^{2}$ Roman HLS survey. $\sim 6$ times difference in number yield from two QLF models. Using QLF from \cite{Matsuoka2018ApJ...869..150M} and \cite{Wang2019ApJ...884...30W}, we predict  of $\sim 180$ quasars in total at $z>6.5$, with $\sim 8$ at $z = 8.0 - 9.0$. }
    \label{fig:predicted_yield_dz_dJ}
\end{figure*}

\begin{deluxetable}{cccc}[h!] 
\tabletypesize{\footnotesize}
\tablenum{2}
\tablecaption{Summary of predicted quasar yields of Roman HLS in redshift bins, determined by integrating the QLFs over the color dropout selection functions. Results from Euclid are based on using both Euclid and Rubin optical data, and assuming \citetalias{Jiang2016ApJ...833..222J} QLF. We present the results from both \citetalias{Jiang2016ApJ...833..222J} and \citetalias{Matsuoka2018ApJ...869..150M} QLF,   representing the optimistic and conservative estimates.}
\label{tab:Predicted_yield}
\tablewidth{0pt}
\tablehead{
\colhead{Redshift range} & \colhead{Roman} & \colhead{Roman} & \colhead{Euclid}\\
& \colhead{Matsuoka+18} & \colhead{Jiang+16} & \colhead{Jiang+16}
}
\startdata
$6.5<z<7.0$ & 158 & 616 & - \\
$7.0<z<7.5$ & 60 & 240 & 204 \\
$7.5<z<8.0$ & 20 & 84 & 45 \\
$8.0<z<8.5$ & 7 & 30 & 16 \\
$8.5<z<9.0$ & 3 & 11 & 7 \\
\enddata
\end{deluxetable}

\subsection{High-redshift Quasar Survey Predictions For Color Selection Method}\label{sec:res_color_selection}

We apply our proposed color selection in Sec \ref{sec:color_dropout_selection_method} to the uniform grid of simulated quasars, and show the color selection function completeness in ($M_\mathrm{1450},z$) space in Fig  \ref{fig:completeness_M1450z_colorcut_z65_90}.
Enforce common $5 \,\sigma$ detection in WFI infrared bands with different $5 \,\sigma$ limiting magnitudes ($J$: 26.95, $F$: 26.2) results in lifting the faintest quasar magnitude limit to $J\sim 26$. 
Color selection successfully recovered $J<25 \;(M_\mathrm{1450}>-22)$ quasars at redshift range $6.5<z<9$, with extended coverage through $J \sim 25.5 \; (\mathrm{S/N} \sim 20)$. 

With color dropout technique,  bright quasars at $J<22$ are completely recovered, while selection function completeness drops rapidly after $J=25$. The main reason of the drop is primarily the common $5 \,\sigma$ S/N ratio requirement in the shallower $F$ band, eliminating lower $S/N$ ($<20$ in $J$) quasars which scatter above the 26.95 detection limit. Besides that, faint quasar selection is also sensitive to $J-H$ and $H-F$ colors. Faint quasars exhibit larger color scatter approaching fainter magnitudes, and subsequently are being missed at $>25.5$. 
To allow more feasible follow up studies, we limit our discussion to $J\leq 25.5$, or $J_\mathrm{lim} = 25.5$, unless specifically noticed. 

The predicted quasar yields per redshift bin and magnitude bin are shown in Fig \ref{fig:predicted_yield_dz_dJ}, and the quasar yield is summarized in Table \ref{tab:Predicted_yield}. 
In the Roman HLS field we show the predictions for both \citetalias{Jiang2016ApJ...833..222J} ($k=-0.72$) and \citetalias{Matsuoka2018ApJ...869..150M} ($k=-0.78$) QLF in 0.5 redshift bins. \citetalias{Jiang2016ApJ...833..222J} QLF is derived from brighter quasars ($M_\mathrm{1450}<-24.$) sample at $z \approx 6.$, and their faint end slope $\displaystyle \alpha=-1.90^{+0.58}_{-0.44}$ is not well constrained; \citetalias{Matsuoka2018ApJ...869..150M} QLF includes more faint quasars $M_\mathrm{1450} \approx -22$ and gives $\displaystyle \alpha=-1.23^{+0.44}_{-0.34}$, however the sample becomes less complete at $M_\mathrm{1450} > -22.5$, which may underestimate the the real number of quasars at the faintest end. 
The two QLF functions represent the optimistic and conservative cases for high-redshift quasar yields, and the $\sim 6$ times differences between these two predictions underlies the need for a deep, wide-field high-redshift quasar survey with high completeness, beyond what the current generation surveys could offer. 

We select the candidates based on color selection from three mock catalogs to estimate the performance of color selection on recovering quasar targets and discriminating against contaminants. The results are scaled to meet the Roman survey area yields. The selected candidate list includes 263 quasars, $1.6 \times 10^4$ MLTs and $2.9 \times 10^6$ galaxies ($43 \%$ ETG).
We define $J$ band $5 \,\sigma$ detected quasars as the underlying real quasars with $J_\mathrm{lim}=25.5$. The number of candidates reduce to 215 quasars out of total 230 quasars, $ 1.2 \times 10^5$ galaxies ($ 38 \%$ ETG) and $1.0 \times 10^4$ MLTs.
Our results show that color selections give exceptionally good completeness, on average $\gtrsim 0.85$ at all redshift ranges. The primary color drop out selections are $z-Y$ and $Y-J$, effectively drop sustainable amount of contaminants. At $z=7.5-8.0$, Ly$\alpha$ break lies between $Y$ and $J$, the completeness significantly depends on the photometric scatter in these two filters, results in overall lower completeness.
We plot the cumulative completeness and efficiency as a function of $J_\mathrm{lim}$, shown in Fig  \ref{fig:completeness_efficiency} and summarize the result in Table \ref{tab:completeness_efficiency_z}. The overall completeness and efficiency are high at bright magnitudes. Efficiency falls rapidly after $J_\mathrm{lim}=23.5$, with only $\sim 0.2$ efficiency at $J_\mathrm{lim}=24$, or finding 1 quasar out of 5 possible candidates.

\subsection{High-redshift Quasar Survey Predictions For Bayesian Modelling Comparison}\label{sec:res_bmc_selection}
\begin{figure}[htbp]
    \centering
    \includegraphics[width=0.48\textwidth]{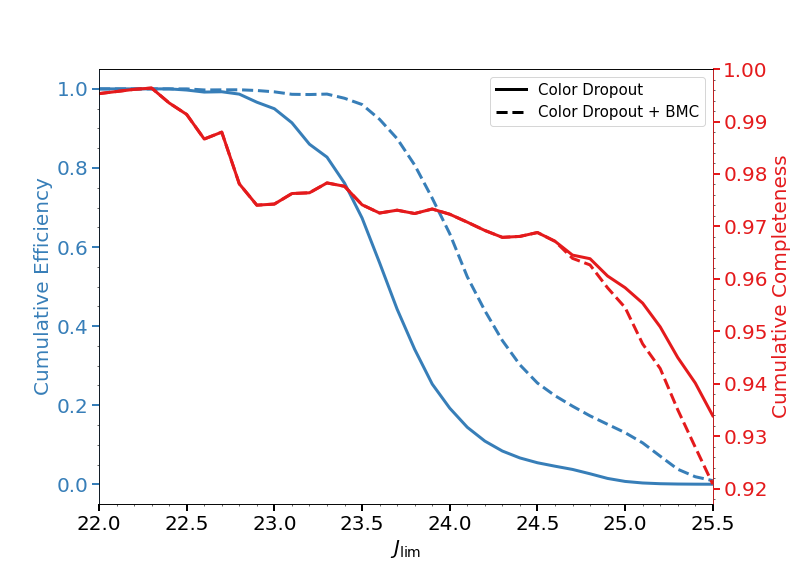}
    \caption{Cumulative efficiency and completeness as a function of $J_\mathrm{lim}$. Overall completeness and efficiency are high at bright magnitude. 
    Optimized selections results in efficiency $\sim$ 0.1 at $J_\mathrm{lim} \sim 25$, around 180 quasars discoverable.  Pushing toward $J_\mathrm{lim}>25.5$, efficiency drop below 0.01, meaning we are expecting one quasar out of several tens to hundreds contaminants.}
    \label{fig:completeness_efficiency}
\end{figure}

BMC is capable of clean selection of bright quasar to $J \sim 23$ at $z=7-9$ \citep{Euclid2019A&A...631A..85E}. 
Color selection gives similar cleanliness results at similar depths. We do not repeat the calculation for bright quasars, and intend to use BMC in improving selection efficiency in the expense of dropping completeness to include fainter quasars. After the color selection, we continue with the Bayesian probability selection. The results are shown in Fig \ref{fig:completeness_efficiency} and Table \ref{tab:completeness_efficiency_z}. 

We adopt a similar Bayesian probability value of  $P_\mathrm{q} = 0.1$ for quasar as in \cite{Mortlock2012MNRAS.419..390M} and \cite{Euclid2019A&A...631A..85E}.
Quasar candidates composes of 211 out of 230 quasars, $2 \times 10^4$ galaxies (40$\%$ ETG), $2.4 \times 10^3$ MLTs at $J_\mathrm{lim}=25.5$. 
We also show cumulative efficiency and completeness down to the faintest $J_\mathrm{lim}$, where selections are still effective. 
Cumulative completeness is quite high even at the faint end, but the cumulative efficiency escalates rapidly after $J_\mathrm{lim}=24$, reaching $0.25/0.12/0.001$ at $J_\mathrm{lim}=24.5/25.0/25.5$, with BMC pushes the identification limit 0.5-1 mag deeper.
Cumulative completeness remain high at the faint end, BMC greatly improve cumulative efficiency at $J_\mathrm{lim}<24$, where similar efficiency level is now 0.5-1 mag deeper than in color dropout selection alone.
We choose a practical 0.1 selection efficiency for sustainable spectroscopic follow up, and conclude that we are able to identify faint quasars at $J \sim 25$ (S/N $\sim$ 15) or $M_\mathrm{1450}\sim -21$. Quantitatively, we are able to recover $\sim$ 180 quasars at $z>6.5$, 23 quasars at $z>7.5$ and 2 quasars at $z>8.5$.  

\begin{deluxetable*}{ccccc}[ht] 
\tabletypesize{\small}
\tablenum{3}
\tablecaption{Summary result of high-$z$ quasar selection functions and quasars recovered from QLF-sampled mock catalogs as described in Sec \ref{sec:mock_catalog_construction} and Sec \ref{sec:high_redshift_quasar_selection}.}
\label{tab:completeness_efficiency_z}
\tablewidth{1pt}
\tablehead{
\colhead{$J_\mathrm{lim}$} &\colhead{Redshift} & \colhead{Color Dropout} & \colhead{Color Dropout+BMC$(P_\mathrm{q} \geq 0.1)$} &\colhead{$N_\mathrm{q}$ Recovered} \\
&& \colhead{
\begin{tabular}{cc}
    C($\%$) &   E($\%$)
\end{tabular} 
} &
\colhead{
\begin{tabular}{cc}
    C($\%$) &   E($\%$)
\end{tabular} 
} &
}
\startdata
24.0 & 
\begin{tabular}{c}
     $6.5<z<7.5$  \\ $7.5<z<8.5$ \\ $8.5<z<9.0$ \\
\end{tabular} &
\begin{tabular}{c}
    \begin{tabular}{cc}
     97.2 & 20.4
    \end{tabular}\\
    \begin{tabular}{cc}
     99.3 & 12.5
    \end{tabular}\\
    \begin{tabular}{cc}
     100.0 & 100.0
    \end{tabular}\\
\end{tabular} &
\begin{tabular}{c}
    \begin{tabular}{cc}
     96.9 & 74.8
    \end{tabular}\\
    \begin{tabular}{cc}
     99.3 & 29.6
    \end{tabular}\\
    \begin{tabular}{cc}
     100.0 & 100.0
    \end{tabular}\\
\end{tabular} &
\begin{tabular}{c}
    45 \\ 7 \\ 1
\end{tabular} \\
\tableline 
& Overall & 
    \begin{tabular}{cc}
     97.5 & 19.0
    \end{tabular} &
    \begin{tabular}{cc}
     97.2 & 63.4
    \end{tabular} & 53 \\
\tableline 
 24.5 & 
\begin{tabular}{c}
     $6.5<z<7.5$  \\ $7.5<z<8.5$ \\ $8.5<z<9.0$ \\
\end{tabular} &
\begin{tabular}{c}
    \begin{tabular}{cc}
     96.8 & 8.7
    \end{tabular}\\
    \begin{tabular}{cc}
     99.0 & 1.5
    \end{tabular}\\
    \begin{tabular}{cc}
     100.0 & 100.0
    \end{tabular}\\
\end{tabular} &
\begin{tabular}{c}
    \begin{tabular}{cc}
     96.5 & 48.2
    \end{tabular}\\
    \begin{tabular}{cc}
     99.0 & 6.0
    \end{tabular}\\
    \begin{tabular}{cc}
     100.0 & 100.0
    \end{tabular}\\
\end{tabular} &
\begin{tabular}{c}
    119 \\ 19 \\ 1
\end{tabular} \\
\tableline 
& Overall & 
    \begin{tabular}{cc}
     97.1 & 5.2
    \end{tabular} &
    \begin{tabular}{cc}
     96.9 & 25.6
    \end{tabular} & 139 \\
\tableline 
25.0 & 
\begin{tabular}{c}
     $6.5<z<7.5$  \\ $7.5<z<8.5$ \\ $8.5<z<9.0$ \\
\end{tabular} &
\begin{tabular}{c}
    \begin{tabular}{cc}
     95.6 & 1.2
    \end{tabular}\\
    \begin{tabular}{cc}
     98.3 & 0.2
    \end{tabular}\\
    \begin{tabular}{cc}
     100.0 & 95.0
    \end{tabular}\\
\end{tabular} &
\begin{tabular}{c}
    \begin{tabular}{cc}
     95.2 & 31.7
    \end{tabular}\\
    \begin{tabular}{cc}
     97.1 & 2.4
    \end{tabular}\\
    \begin{tabular}{cc}
     100.0 & 95.0
    \end{tabular}\\
\end{tabular} &
\begin{tabular}{c}
    151 \\ 23 \\ 2
\end{tabular} \\
\tableline 
& Overall & 
    \begin{tabular}{cc}
     96.0 & 0.7
    \end{tabular} &
    \begin{tabular}{cc}
     95.5 & 13.1
    \end{tabular} & 176 \\
\tableline 
25.5 & 
\begin{tabular}{c}
     $6.5<z<7.5$  \\ $7.5<z<8.5$ \\ $8.5<z<9.0$ \\
\end{tabular} &
\begin{tabular}{c}
    \begin{tabular}{cc}
     93.4 & 0.03
    \end{tabular}\\
    \begin{tabular}{cc}
     93.8 & 0.007
    \end{tabular}\\
    \begin{tabular}{cc}
     95.7 & 0.2
    \end{tabular}\\
\end{tabular} &
\begin{tabular}{c}
    \begin{tabular}{cc}
     92.5 & 1.7
    \end{tabular}\\
    \begin{tabular}{cc}
     88.9 & 0.2
    \end{tabular}\\
    \begin{tabular}{cc}
     95.7 & 1.2
    \end{tabular}\\
\end{tabular} &
\begin{tabular}{c}
    182 \\ 27 \\ 2
\end{tabular} \\
\tableline 
& Overall & 
    \begin{tabular}{cc}
     93.4 & 0.02
    \end{tabular} &
    \begin{tabular}{cc}
     92.1 & 0.9
    \end{tabular} & 211 \\
\tableline 
\enddata
\end{deluxetable*}

\subsection{Summary on Joint Selection Functions}
At brighter magnitude $J<23$, color drop out and color cut effectively remove majority of the contaminants. Similar completeness and efficiency at bright end has been reported in \cite{Euclid2019A&A...631A..85E}, as a result we focus on the performance of BMC at faint end.
Toward fainter magnitude, Bayesian probability method improves the  efficiency by a factor of ten, while retains similar completeness. We examine the performance of BMC by adjusting $P_\mathrm{q}=0.01-0.9$, with trade off of worse completeness for exchange of higher efficiency. The results are insensitive to the choice except for extreme threshold values. We decide to keep $P_\mathrm{q}=0.1$ as a fair comparison to literature studies. 

To investigate the effect of selection functions optimized in both identifying quasars and rejecting contaminants, we display the number of recovered quasars as a function of $J_\mathrm{lim}$ magnitude, shown in Table \ref{tab:completeness_efficiency_z}. 
At $J \sim 25$, we are expecting to find at least 1 $z>8.5$ quasar simply by color dropout given HLS 2000 deg$^{2}$ survey area. At $J<24.5$, galaxies are less of concern since they can be generally eliminated by purely color dropout, MLTs are the major contaminants because of color similarity. 
Moving toward fainter magnitudes, unresolved galaxies of high surface densities and low S/N photometry begin to dominate contamination. We note that galaxies interfere more at $z=6.5-7.5$ than $z=7.5-8.5$ color dropout selection. Inclusion of BMC effectively remove galaxies from candidates at $J_\mathrm{lim}=25$.
MLTs are the main contaminant in this study at brighter magnitude. They are extremely difficult to remove based on our current models, especially for spectral types L2-L4 and T2-T4. 
These cool dwarfs have low-temperature atmospheres where their red infrared colors are dominated by strong $\mathrm{H_2O}, \mathrm{CH_4}$ absorption features that interfere with the $J-H$ and $H-F$, and are indistinguishable from quasar colors. The inclusion of the surface density prior does remove some MLTs from the contamination, but additional parameters are needed to distinguish faint MLTs from quasars, i.e. the proper motion, which will be discussed in the following Sec \ref{sec:proper_motion}.

\subsection{Highest Redshift Quasar in the Universe}
Quasar spatial density at $z>6$ has shown to decline more rapidly than at lower redshifts \citep{Richards2006AJ....131.2766R,McGreer2013ApJ...768..105M, Jiang2016ApJ...833..222J,Wang2019ApJ...884...30W}. \cite{Wang2019ApJ...884...30W} predicts $\sim$ 1 luminous quasar available over all sky region at $z>9$ with $J\sim 21$. 
We investigate how to maximize the probability of finding the highest redshift quasars, $\sim 5-7$ quasars with $J_\mathrm{lim}=25.5-26.95$, discoverable by Roman.
At $z>8.4$, the quasar Ly$\alpha$ break redshifts into $J$. Candidates can be selected through dropping out all bluer bands. 
We require candidates exhibits non-detection in $u,g,r,i,z$. We use $Y-J>1.5$ if detected, or S/N$(Y)<3$ if undetected, to select potential $z>8.4$ quasar candidates. 
We apply the above color selection on the mock populations at $J_\mathrm{lim}=25/25.5/26$. The resulting candidates include 3/4/5 out of 4/5/6 quasars, $25 / 1.7 \times 10^4 / 6.7 \times 10^5$ galaxies ($64/63/61 \%$ ETG), $1 / 20 / 680$ MLTs, cumulative completeness $\geq 0.75$ and cumulative efficiency $0.12/2.5\times 10^{-4} / 7.5\times 10^{-6}$. The efficiency declines rapidly after $J_\mathrm{lim}=25$ or S/N=30. We further investigate if detection and color cut in $H,F$ would improve our selection. We find that using 1. S/N$(H,F)>5 \,\sigma$ 2. $J-H>0.1$ and 3. $-0.2<H-F<0.1$ result in slightly better efficiency at $J_\mathrm{lim}=25.5$, where there are 4 quasars, $1.6\times 10^3$ galaxies ($61\%$ ETG) and 2 MLTs. We summarized the color selection as below, 

{\footnotesize
\begin{align}\label{eq:highestz color dropout}
    &\mathrm{S/N} (u,g,r,i,z)<5.0  \nonumber \\
    &\mathrm{S/N}(J,H,F)>5.0 \nonumber \\
    \text{For $z>8.4$:}&
    \begin{cases}
     \mathrm{S/N}(Y)<3.0 \text{ or } Y-J > 1.5 \\
     J-H > 0.1 \\
     -0.2 <H-F < 0.2 \\
    \end{cases}
\end{align}}%

Color selection alone enables high completeness selection of highest redshift quasars down to $J_\mathrm{lim}=25.$ with $0.1$ efficiency.
We further proceed with $P_\mathrm{q} = 0.1$ selection. At $J_\mathrm{lim}=25.5$, there are 4 out of 5 quasars, 47 galaxies ($54\%$ ETG) and 1 MLT, selection efficiency is around $0.1$ but 0.4 mag deeper than in Sec \ref{sec:res_bmc_selection}. Number of $z>8.4$ quasars have doubled, with the highest possible redshift at $z \sim 9.6$ with $J\sim 25$. Finding these earliest and faintest quasars is important because they may host the first generation of SMBHs and are critical for understanding SMBH growth and SMBH-galaxy coevolution in their earliest stages.

\section{Discussions} \label{sec:discussion}
\subsection{Proper Motion}\label{sec:proper_motion}
To further reduce the amount of faint MLTs contaminants, we investigate the possibility to improve the selection by considering proper motion based on the study in \cite{WFIRST2019JATIS...5d4005W}. 
We find that Rubin LSST ability to identify the cool brown dwarfs through proper motion is limited for our application, because the majority MLTs that we are concerned are at $J \gtrsim 23.5 \,(500 \;\mathrm{pc})$. These targets are largely undetected or marginally detected in the LSST bands, with poor astrometry. When LSST completes 5 years survey, Roman will have completed 1 year mapping on the HLS region, by using forced photometry, it may be able to identify moving objects and place constraint on the position offsets. Dwarf thin disk population with $v_\mathrm{tan} = 30 \;\mathrm{kms}^{-1}$ move with $\mu \sim 10 \mathrm{\;mas \; year}^{-1}$, which is significantly smaller than Rubin LSST limit $\mu_\mathrm{LSST} = 35 \mathrm{\;mas \; year}^{-1}$, and $< 5\%$ of original MLT population can be rejected.

\begin{figure}[h]
    \centering
    \makebox[\linewidth][c]{\includegraphics[width = 1. \linewidth]{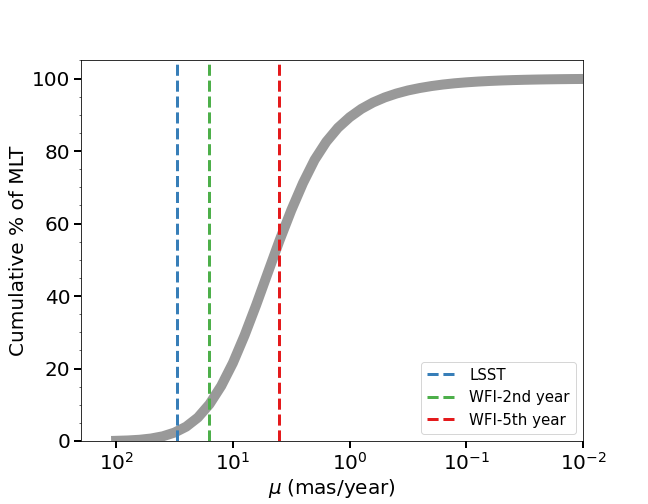}}%
    \caption{Cumulative percentage of the MLTs population as a function of proper motion. LSST can observe cool brown dwarf with $\mu_\mathrm{LSST} = 30 \mathrm{\;mas \; year}^{-1}$. WFI has the ability to measure $\mu_\mathrm{WFI} = 4-16 \mathrm{\;mas \; year}^{-1}$. The curve represents the percentage of MLTs with proper motion greater than $\mu$, dashed lines represent the proper motion limits of each survey. 
    }
    \label{fig:proper_motion_mlt_cum_percentage}
\end{figure}

The exact cadence of Roman/HLS is not yet finalized. To investigate the possibility of using proper motion obtained from Roman/HLS, 
we assume a second scan of the same survey area after a time lag of 1 year and 4 years, with $5 \,\sigma$ requirement in localization, correspond to $\mu_\mathrm{WFI} = 16 \mathrm{\;mas \; year}^{-1}$ and $4 \mathrm{\;mas \; year}^{-1}$, resulting in $6 \%$ and $46 \%$ removal of targets. 
The result is shown in Fig \ref{fig:proper_motion_mlt_cum_percentage}. 
In the most optimistic scenario, we expect $50\%$ MLT population can be rejected from quasar candidates.
Follow results from joint selection functions, assuming optimistic $\mu_{\mathrm{WFI}}=4 \mathrm{\;mas \; year}^{-1}$ threshold, improvement on selection efficiency are $70\%$ at $J_\mathrm{lim}=25$ and $6\%$ at $J_\mathrm{lim}=25.5$. 
Proper motion measurements improve the selection efficiency at brighter magnitude regime ($J \sim 24$), the improvement is only marginal at the faintest magnitudes, where contamination is dominated by unresolved galaxies rather than MLT dwarfs.

\subsection{Constraint on QLF Faint End Slope}
Bridging the high-redshift faintest quasars/AGNs and brightest galaxies can provide understanding on the coevolution between AGNs and their host galaxies. 
Recent discussions have been focusing on transition at $M_\mathrm{UV}\sim -22$ \citep[e.g.][]{Kulkarni2019MNRAS.488.1035K,Harikane2022ApJS..259...20H,Adams2022arXiv220709342A} where AGNs LF appears to flatten. 
\citetalias{Matsuoka2018ApJ...869..150M} use 18 $M_\mathrm{1450}>-23.25$ low luminosity quasars for faint end slope determination.  \cite{Shen2020MNRAS.495.3252S} and \cite{Finkelstein2022arXiv220702233F} compile all known quasars and galaxies with rest-frame UV detection at $z \sim 6$, their models favor a shallow faint end slope for AGNs $M_\mathrm{UV}>-23$, although they do not completely rule out the steeper slope due to limited sample sizes. With Roman/Rubin capabilities in finding $J_\mathrm{lim}=25$ quasars at $z>6.5$, we are going to discover $>80$ at $M_\mathrm{1450}>-23$, 30 at $M_\mathrm{1450}>-22$. With 4 times increase in low luminosity quasar sample sizes, the faint end slope can be better constrained with 2 times smaller in uncertainties, we are able to distinguish between \citetalias{Jiang2016ApJ...833..222J} and \citetalias{Matsuoka2018ApJ...869..150M}, finally determining the underlying faint population density. 

\section{Summary}\label{sec:summary}
In this study we predict quasar yields at $z>6.5$ in the era of Roman and Rubin.
We incorporate the public information from Roman and Rubin, including optical/IR filter wavelengths and limited to Roman 2000 deg$^{2}$ HLS survey area. The results are summarized below.
\begin{enumerate}
    \item With Roman and Rubin imaging capabilities, using joint selection of color and Bayesian Model Comparison, $\sim 180/200$ quasars with $J_\mathrm{lim}=25.0/25.5$ will be discovered at $6.5<z<9.0$ within 2000 deg$^{2}$ assuming \citetalias{Matsuoka2018ApJ...869..150M} and \citetalias{Wang2019ApJ...884...30W} prescription, expanding at least 3 times the current high-$z$ quasar samples. The estimated cumulative completeness is $\geq 0.8$, while the cumulative efficiency drops below 0.1 after $J_\mathrm{lim}=25$.
    \item We expect to find around 4 quasars at $z>8.4$, with the possible highest redshift quasar at $z>9.5$ with $J \sim 25$.
    \item QLF plays important role in determining the predicted high-$z$ quasar yields. Using the conservative \citepalias{Matsuoka2018ApJ...869..150M,Wang2019ApJ...884...30W} and optimistic \citepalias{Jiang2016ApJ...833..222J} QLF parameters result in $\sim 6$ times difference in the quasar yields. With Roman/Rubin capability in finding a statistical faint quasar sample, we can put stronger constrain on the QLF faint end slope.
    \item Color selection alone is sufficient to select $J_\mathrm{lim}<24$ quasars with significant Ly$\alpha$ break. BMC with surface density prior can push the selection threshold 1 mag deeper. 
    \item Contamination from low-$z$ sources, mainly low S/N unresolved galaxies and cool $LT$ dwarfs, is prominent at $J>24$. Selection efficiency at brighter magnitudes can be improved through proper motion detection of galactic dwarfs. The improvement is marginal at faintest magnitudes where low-$z$ galaxies dominates contamination.
\end{enumerate}

We emphasize that current deployment time of Roman and Euclid is still uncertain. Predictions in \cite{Euclid2019A&A...631A..85E} and this work provide an estimated range for the number of the earliest high-$z$ quasars across wide range of luminosities discoverable in the next decade. 
This work focuses on exploring the earliest and faintest quasar population, which are essential part in constraining the high-$z$ supermassive black hole abundance and understanding their growth in the early Universe.

\section{Acknowledgement}\label{sec:acknowledgement}
WLT appreciate the comments and suggestions from R. Green 
for useful and informative discussion. 
We thank the community feedback from WFIRST Science Investigation Team (SIT) supported via NASA contract NNG16PJ33C 'Studying Cosmic Dawn with WFIRST', Early Universe/Reionization Era Conversations at Arizona and Roman Science Team Community.
WLT and XF acknowledge supports by NSF grants AST 19-08284. 
FW thanks the support provided by NASA through the NASA Hubble Fellowship grant $\#$HST-HF2-51448.001-A awarded by the Space Telescope Science Institute, which is operated by the Association of Universities for Research in Astronomy, Incorporated, under NASA contract NAS526555. 
WLT would like to thank the support by Lia YCC throughout the work.

\textit{Software}: IPython \citep{ipythonPER-GRA:2007}, matplotlib \citep{matplotlibHunter:2007}, NumPy \citep{numpyharris2020array}, SciPy \citep{scipy2020SciPy-NMeth}, Astropy \citep{astropy:2013,astropy:2018,astropy:2022}, seaborn \citep{seabornWaskom2021}, GCRCatalogs \citep{GCRCatalogsMao2018}, SIMQSO \citep{McGreer2013ApJ...768..105M}, Pq\textunderscore server \citep{Barnett2021MNRAS.501.1663B}.

\bibliography{highz_quasar}
\bibliographystyle{aasjournal}



\end{document}